\documentclass[runningheads]{llncs}

\usepackage[T1]{fontenc}

\usepackage{graphicx}
\usepackage{caption}
\usepackage{subcaption}
\usepackage{url}

\begin{document}

\title{Space-efficient B-tree Implementation for Memory-Constrained Flash Embedded Devices}
\titlerunning{Space-efficient B-tree Implementation for Embedded Flash Devices}

\author{Nadir Ould-Khessal \and
Scott Fazackerley \and
Ramon Lawrence}

\institute{University of British Columbia}

\maketitle            

\begin{abstract}

Small devices collecting data for agricultural, environmental, and industrial monitoring enable Internet of Things (IoT) applications. Given their critical role in data collection, there is a need for optimizations to improve on-device data processing. Edge device computing allows processing of the data closer to where it is collected and reduces the amount of network transmissions. The B-tree has been optimized for flash storage on servers and solid-state drives, but these optimizations often require hardware and memory resources not available on embedded devices. The contribution of this work is the development and experimental evaluation of multiple variants for B-trees on memory-constrained embedded devices. Experimental results demonstrate that even the smallest devices can perform efficient B-tree indexing, and there is a significant performance advantage for using storage-specific optimizations.


\end{abstract}

\section{Introduction}

Devices deployed at the edge of networks are collecting and processing large amounts of data. Processing data on the device before sending it to servers may improve response time and energy efficiency and reduce the amount of data transmitted over the network. The smallest embedded devices have unique hardware and performance characteristics that challenge data processing. These devices may have very small memory between 4 to 64 KB, limited CPUs, and use raw flash storage without a file system. Data processing is dominated by inserts, and there is a limited number of queries related to the collected data.

This work optimizes the B-tree \cite{btree} for this environment. Prior research adapted the B-tree to flash-based storage such as solid-state drives (SSDs). A survey of flash indexing \cite{indexsurvey} overviews over 30 implementations. A common approach is to reduce the number of expensive flash writes by using available memory to defer writes and allow for batching and sequential writes. Systems use either a logging approach \cite{lsbtree,iplbtree,asbtree,bftl} or buffer page updates before writing \cite{betree}. The highest-performing algorithms exploit the inherent parallelism in SSDs \cite{indexsurvey}.

A shorter version of this work appeared in CoopIS 2024 \cite{coopis24}. This paper extends the conference version with expanded background on flash indexing techniques, a more detailed description of the virtual mapping and recovery mechanisms, additional discussion of storage management strategies, and a more comprehensive presentation of the experimental evaluation.

Implementing a B-tree for a memory-constrained embedded device has unique challenges compared to indexing on servers. The most significant issue is the limited memory. This requires an algorithm to consider its total memory consumption beyond the memory used for buffers. Memory used for implementation and state variables is important. Although the data sizes processed are orders of magnitude smaller, the relative RAM versus storage ratio is lower. Typical servers have at least 1\% RAM compared to the data storage size, and many deployments have enough memory to store a large percentage of the data set \cite{graefe5min}. An embedded device may have less than 1\% memory compared to the data size, and the absolute memory size is very small. Every byte counts in this environment.

The second factor is the flash storage may consist of raw flash memory chips rather than packaged products like SD cards or solid-state drives. This affects implementations in two important ways. First, the implementation must handle the management of physical memory on the devices including wear leveling, page placement, and free space management. Without a flash translation layer (FTL) \cite{ftl}, the algorithm must map logical pages to physical pages while using minimal memory. Secondly, there are fewer opportunities for parallelism. The CPU is usually single-threaded, the storage device can only respond to one request at a time, and transferring data on the bus is a bottleneck. I/O performance is limited by the bus speed rather than the storage device bandwidth. 

This work proposes and evaluates optimizations and variants for B-trees in the memory-constrained, embedded environment. The contributions include:

\begin{itemize}
    
    \item An approach to reducing write amplification for B-tree updates by using virtual mappings. Virtual mappings allow physical page movements without rebuilding the B-tree index structure.

    \item A B-tree variant utilizing virtual mappings for raw NAND flash storage. 
      
    \item A B-tree variant designed for memory supporting page overwriting such as NOR and DataFlash \cite{dataflash}. 
    
    \item An experimental evaluation on two hardware platforms and three memory types showing the benefits of storage-specific B-tree optimizations to maximize performance on hardware with unique characteristics.
    
    \item Analysis on the most effective usage for memory for the B-tree index when the memory available is small.
    
\end{itemize}

The organization of the paper is as follows. Section 2 provides background on the B-tree data structure and implementations for flash-based indexing. Section~3 provides a motivating environment for the work. Section 4 discusses optimizations for B-tree implementations with focus on new techniques for managing virtual mappings and page overwriting. Section 5 has experimental results. The paper closes with future work and conclusions.

\section{Background}

\begin{figure*}[!thbp]
	\centering
	\includegraphics[width=4.7in]{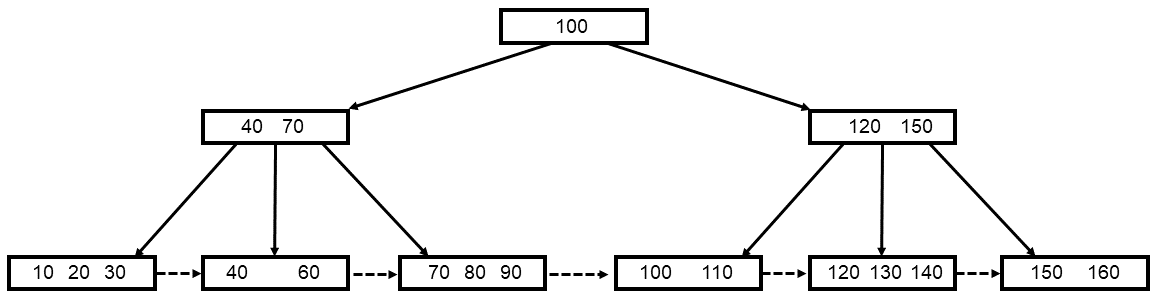}
	\caption{B\textsuperscript{+}-tree Example}
	\label{btree}
\end{figure*}

The B-tree \cite{btree} is a high-performance index structure for database applications. Most implementations use the B\textsuperscript{+}-tree variant \cite{comer79} where data records are only in the leaf nodes, and interior nodes contain keys for navigation. An example B\textsuperscript{+}-tree is in Figure \ref{btree}, where each node contains a maximum of 3 entries.

There is a long history of B-tree implementations with an initial focus on hard drives. Graefe \cite{graefebtreelogging} provides a detailed discussion of numerous challenges, techniques, and best practices for B-tree implementation. Foster B-trees \cite{fosterbtree} combine the advantages of several B-tree variants including the write-optimized B-tree \cite{writeopttree} to support improved concurrency and efficient page migration in flash storage. Foster B-trees eliminate sibling pointers in the B-tree structure (dashed pointers in the figure) and rely only on parent-children pointers. 

Virtual memory and shadow paging techniques \cite{shadowpaging} have been applied to B-trees \cite{gray93,btreeshadow}. Bw-trees \cite{bwtree} use a mapping table to convert logical pointer addresses to physical page locations. This allows a page to be physically written to any storage location without requiring updating pointers on the complete path from leaf to root. The performance of the mapping table is critical. 

The migration from hard drives to flash-based storage and solid-state drives (SSDs) resulted in a renewed emphasis on B-trees and optimization for the flash characteristics \cite{indexsurvey}. The key benefit of flash is the significantly higher random access performance compared to the high seek times for hard drives. This motivates having B-tree nodes of smaller size when storing on SSDs compared to hard drives. A challenge is handling the flash memory property that in most cases a physical page cannot be overwritten in-place without first performing a block erase. Writes are often more costly than reads leading to asymetric performance. 

B-tree implementations adapted to flash memory properties by favoring reads over writes, using in-memory buffers and logging to reduce the number of writes, and deferring and batching random writes into sequential writes. An SSD has an internal CPU and memory and can process multiple requests simultaneously. As SSD technology has evolved, the performance between reading and writing has been reduced by the internal hardware in the SSD that uses its flash translation layer (FTL) algorithm to intelligently manage physical memory pages. 

To understand a fundamental challenge, consider inserting key 50 into the B-tree in Figure \ref{btree}. Key 50 will be inserted into the second leaf node currently containing 40 and 60. On a hard drive, the page containing the node is updated and written back in-place on disk. No modifications to the B-tree structure are required. On an SSD, the B-tree implementation may also overwrite the page logically, but the SSD must use its FTL to write the page to a different physical location as overwriting the same location without erasing is not supported. 

For raw flash memory chips that do not have an FTL and file interface, the B-tree implementation must change. It is not possible to physically overwrite an updated page, and writing to a new page location requires updating B-tree pointers. Updating a pointer may require cascading updates to other pages. In our example, inserting 50 and writing the updated page to a new location requires updating the pointer in the parent node containing (40,70) that then triggers an update and write of the root node. Further, if sibling pointers are used, the leaf node containing (10,20,30) would also require updating. Reducing this write amplification is critical to ensure acceptable performance.

A survey on flash based indexing \cite{indexsurvey} overviews various B-tree implementations. Buffering approaches such as write-optimized trees \cite{betree} defer writes by buffering nodes in memory and writing in batches to amortize the write cost and avoid small random writes. Logging approaches \cite{lsbtree,iplbtree,asbtree,bftl} log changes in memory or on the flash data page rather than updating immediately. In our insert key 50 example, a buffering approach may buffer the leaf node and its parent in memory and only write the nodes after many updates have occurred. A logging approach would store the insert 50 as a log record in a separate memory area and defer making changes to the tree. The tree is updated in batch when the log area is full. Using memory for buffering or logging improves performance by maximizing parallel usage of SSD hardware resources and capabilities.  

Raw flash B-tree implementations often modify the B-tree node structure to reduce the write amplification. These implementations manage memory including free space and garbage collection. Some implementations exploit flash memory properties such as partial page updates or overwrites \cite{btreeoverwrite2,btreeoverwrite1}. B-tree implementations for flash memory and SSDs based on buffering and deferred writes are not executable on memory-constrained embedded devices if they do not adapt to the low memory available. A B-tree implementation \cite{btreeembedded} for embedded devices that uses limited memory is available but is not optimized for raw flash and relies on file-based storage.

Other index structures for embedded devices include Antelope \cite{dbeverysensor}, MicroHash \cite{microhash}, PBFilter~\cite{pbfilter}, and SBITS \cite{sbits}. These index structures, typically for sequential time series data, use sequential writes and minimize memory usage. They do not support indexing for non-sequential data like a B-tree.

\section{Motivating Environment}

The target environment is resource-constrained embedded devices primarily used in logging and monitoring applications. The device may have a CPU speed of 16 MHz to 128 MHz, between 4 KB and 64 KB RAM, and use NOR, NAND, or SD card storage. The flash storage may not have an FTL or file system. The flash storage size depends on the data collection requirements and may be between 8 MB to 1 GB. These hardware specifications (see Figure \ref{devices}) are comparable to the IBM 360 machine used in 1970 for the original B-tree experiments.

\begin{figure}[!ht]
	\centering
	\begin{tabular}{|l|c|c|c|}
	\hline
	{\bf Device} & {\bf CPU} & {\bf RAM}  & {\bf I/O Rate} \\
	\hline
    M0+ SAMD21  & 48 MHz & 32 KB  & 450 KB/s \\
    PIC24FJ1024 
                & 16 MHz & 32 KB & 200 KB/s \\
    IBM 360/44 \cite{btree} & 1.6 MHz & 32-256 KB & 156 KB/s \\
	\hline
	\end{tabular}
	\caption{Embedded Devices and Original Experimental B-tree Hardware in 1970}
	\label{devices}   	
\end{figure}


A page is the I/O unit for flash storage and has size between 256 and 4096 bytes. A page must be erased before it is written. Erasing is done in units of blocks, which are consecutive pages. Writes are not performed in-place due to the cost of erasing a block. If SD card storage is used, the flash translation layer (FTL) writes to a new physical location and updates its internal logical to physical page mapping. When using raw memory with no FTL, the algorithm must manage physical page allocation, garbage collection, and wear leveling. Some NOR flash memory supports limited page overwriting \cite{dataflash}. It is possible to overwrite a page as long as no bits are changed from 0 to 1. Bit changes from 1 to 0 are acceptable. This is achievable by initializing a page to all ones and appending to free space in the page that has not been previously written. 

The logging environment domain is characterized by very constrained and specific data access requirements. The devices are typically collecting time series data consisting of a timestamp and a sensed value which is often a small number. Highly efficient inserts are the key consideration. Record inserts are typically append-only in order of record timestamp. Updates and deletes are very rare. Bulk deletes may occur periodically when the device eliminates historical data. For example, older data may be deleted from the system on a daily basis after it is no longer useful or has been transmitted to a server for processing. Query processing on-device is for performing searches by timestamp or value and aggregate queries over time windows. These queries are generated by the device as part of its functions and are generally known before deployment. The low RAM challenges data processing, and minimizing memory usage is critical. The RAM to storage ratio is lower than server systems \cite{graefe5min}. For example, if 4 KB RAM is available, that is 1\% of a 400 KB data size and 0.1\% of a 4 MB data size. 

The B-tree implementation should adapt to these different device properties. There are two distinct use cases for embedded indexing. Indexing the append-only, sorted time series data is important for fast retrieval based on timestamp. Numerous prior index structures that utilize sequential writes, including simple sorted files, demonstrate good performance with minimal memory usage \cite{sbits,indexsurvey,dbeverysensor,pbfilter,microhash}. Even when optimized for sequential data, the B-tree has comparable, but not superior, performance to these specialized index structures \cite{embedindexsurvey}.

Retrieval by data value in the time series is not as efficiently handled by current embedded index structures. There is no ordering of the data values, and the index must handle random writes to any location. There may also be a considerable amount of duplication in the data set. Further, if the sensor is capturing physical measurements of the environment, the data may be more periodic and follow a function based on the value being measured. As an example, temperature values change slowly over time in a relatively predictable fashion. The focus of this work is on indexing non-sorted data efficiently.

\section{B-tree Optimizations}

This section presents B-tree optimizations and implementation variants for embedded systems. Previous work on logging and write-optimized trees \cite{betree,lsbtree,iplbtree,asbtree,bftl} are combined and compared with an approach that uses virtual mappings to defer writing nodes when their children pointers change. Although previous B-tree implementations may not be directly executable in this environment as they either consume too much memory or rely on operating system features that are not available, the key optimizations are transferable and adaptable.

\subsection{Buffering and Logging}

Many B-tree approaches optimized for flash memory rely on either buffering updates to defer writes or logging changes to later perform the operations in batch. Adapting these techniques to the embedded environment requires handling that the memory available is small. There is often only sufficient memory to buffer the root node and a few nodes in the working path, and not enough to buffer even a small fraction of the children nodes of the root. An approach that updates a B-tree node directly in the buffer and defers writing to storage has minimal performance benefit as a page will not remain in the buffer for long enough to accumulate multiple updates before being written to storage. 

Logging operations to perform in batch is more memory efficient as the operation is stored rather than the modified page. This allows for more modifications in the limited memory. The diversity of previous log-based approaches associated logs with either the entire tree, individual nodes, or based on storage blocks. The most memory efficient is to use a single log for the tree that functions as a write buffer. When the write buffer is full, the operations are sorted, and the inserts are performed in batch. This reduces I/Os by inserting multiple records into a page at a time and has good performance for clustered data.

The write buffer optimization performing batch inserts can be applied with any of the other B-tree optimizations. For applications with restrictions on data loss, the write buffer can be sized to require operations to be performed after a given amount of time or records have been accumulated.

\subsection{Virtual Mappings}

Storage devices such as SSDs and SD cards support logical page overwriting as the FTL translates logical writes of pages in a file to physical memory locations on the device. At the hardware level, the storage device must translate logical page requests to physical locations and maintain information on pages that are no longer valid. Transparent to the application, the device erases blocks containing invalid pages and maintains its storage allocations. Since it is not possible to update in-place on raw flash storage, updating a node results in writing to a new physical page, and B-tree pointers to the previous physical page must be updated. Updating pointers cascades to many nodes in the tree, and writing a leaf node may result in writing the entire leaf-to-root.

Virtual mappings resolve the write amplification issue when writing B-tree nodes on flash storage without an FTL. Since it is not possible to implement an FTL in the limited device memory, virtual mappings are a B-tree specific translation mechanism. A virtual mapping is a pair consisting of its previous and new page id {\em prevPageId} $\rightarrow$ {\em newPageId}. Whenever a page is written to a new location, a virtual mapping is inserted into the mapping table. A page id is mapped to a physical address based on the starting memory address and multiplying the id by the page size to use as an offset. In the example when inserting key 50, the node containing (40, 60) is updated and written to a new location. The parent node containing (40, 70) requires a pointer update to point to the new location. Rather than modifying the parent node, a virtual mapping is created. Whenever the pointer is traversed, the mapping table is used to lookup if a new location is used.

To minimize the number of mappings, similar to Foster B-trees \cite{fosterbtree}, there are no sibling pointers. With only parent pointers, there is at most one pointer to any given node.
Virtual mappings are similar to the mapping table used in Bw-trees \cite{bwtree} except that due to the limited memory, the mapping table size is bounded, and only the required mappings are in the table. Given a page id as a pointer in a B-tree node, if no mapping is found in the table, then the page id provides the node's current physical location. Thus, the mappings are more similar to physical redirects rather than logical to physical mappings. A mapping requires no additional I/Os and will not be chained multiple times. Consider if the node containing (40, 60) is written multiple times at different locations $L_1, L_2, ..., L_N$ and the parent node (40, 70) pointer stores the pointer value $L_1$. On writing to location $L_2$, the mapping $L_1 \rightarrow L_2$ is added to the table. Each following write only updates the mapping $L_1 \rightarrow L_3$, $L_1 \rightarrow  L_4$, $...$, to finally $L_1 \rightarrow  L_N$. There is no need to store any intermediate mappings as there is only one node that uses the mapping, and it contains the value $L_1$. Before any I/O is performed, the mapping table is searched to determine if a page's physical location has changed. In all cases, only one I/O is required to retrieve the page.

Virtual mappings must be memory-resident. Otherwise, any pointer lookup requires an I/O to determine if a mapping exists. A virtual mapping record size is 4 or 8 bytes depending on the number of storage pages. The mapping table size is configurable but is typically only about 1 to 2 KB. Our implementation is tuned to minimize memory while supporting fast lookups and uses an in-memory hash table with double hashing to handle collisions. Mappings can be removed from the table either during an insert operation or in-between B-tree operations. During an insert operation, the number of mappings on each node on the path to the leaf can be determined, and the node written to storage if the mappings exceed a certain amount. There is only a net benefit in the number of mappings if a node is written that has 2 or more mappings. This is because whenever a node is updated its mappings can be removed (as its pointers are updated), but its parent will need a new mapping for its new location. At any time, a traversal from the root can be performed to identify nodes with mappings that can be removed. Writing a full path from leaf-to-root removes all mappings along that path. If there is no space to store a mapping, then the node that required the mapping is written to storage. With sufficient space to store all mappings, there is no write amplification. Using virtual mappings allows sequentially writing pages to storage. When a page is updated, it is written to the next sequential page location, and a virtual mapping created. This has performance benefits for some devices, and also simplifies free space management and garbage collection.

\subsection{Page Overwriting}

For raw memory storage, the application maintains the physical memory including free space management and block erases. Typically, a page that has been previously written cannot be overwritten in-place. However, there are some memory devices that allow limited page overwrites which can be exploited to improve performance \cite{dataflash}. Specifically, some NAND flash allow partial page writes and NOR and DataFlash products support overwriting of a complete page given constraints. Due to the nature of NOR flash, the entire page can be overwritten but only bit changes from 1 to 0 are supported \cite{dataflash,Overwrite_UBC}. Practically, this allows writing to parts of the page that were not written to before.

Overwriting improves write performance and reduces the number of erases. In \cite{btreeoverwrite2}, the B-tree node is modified to store byte-level differences. Our implementation stores the B-tree node in linear insertion order (i.e. not sorted) which works well with overwriting. Each record is appended to the page. There are two bits associated with each record: a count bit and a valid bit. Both bit vectors are initialized to 1s. When a record is inserted into the page, the count bit is set to 0 from 1. The valid bit remains as 1. If a record is moved or deleted, its valid bit is set to 0. The number of valid records in a page is determined by adding up all records with a count bit of 0 and a valid bit of 1. This structure allows for all updates to have writes of 1 to 0 as supported by the memory. Since keys are no longer sorted, linear search is used.

\subsection{Free Space Management}

Free space management and garbage collection is implemented for storage that does not provide these functionalities. The storage is treated as a circular array. Several blocks are erased preceding the current write location at all times. Before erasing a block, valid pages are determined and buffered in memory or storage. Once a block is erased, the valid pages are written back to their original locations in the block so no mappings must be updated in the tree.

To determine if a page is valid there are two approaches. The first approach associates a free space bit with each page. Once a page is no longer used, set the bit indicating it can be recycled. If there is insufficient memory for a bit vector, a page search strategy can be used to determine if the page is valid. The algorithm to determine that any given page on storage is a valid page is to read the page $P$ from storage, select a key $K$ from the page, and then search the current tree for $K$. If $K$ is not found or a different page encountered, then $P$ is not a valid page and can be deleted.

\subsection{Recovery}

Every page header stores a page id and a previous page id. The previous page id is -1 for a new page written for the first time. When writing a new version of an existing page with ({\em pageId}, {\em prevId}), the previous page id for the new page is {\em prevId} if there exists a mapping {\em prevId} $\rightarrow$ {\em pageId} else it is {\em pageId}. This allows tracking the provenance of a page and recovering the virtual mapping table if there is a device failure or restart.

On restart, the storage is scanned backwards from the last page write. During this scan, the most recent root node is identified by the page that has the root flag set in the header and has the largest page id. When a page is read with header ({\em pageId}, {\em prevId}), search the mapping table for {\em prevId}. If no mapping is found, a mapping is added to the mapping table of the form {\em prevId} $\rightarrow$ {\em pageId}. If a mapping {\em prevId} $\rightarrow$ {\em pageId2} is found and {\em pageId2} $>$ {\em pageId}, then the {\em pageId} page is an older version and is marked as a free page in the free space bitmap.

Consider a leaf node at location 5 written three times due to inserting three keys. The first page has a header with ({\em pageId}, {\em prevId}) of $(5,-1)$. The second page has header $(6,5)$. The third page has header $(7,5)$ and the fourth page has header $(8,5)$. Scanning the storage backwards from the last page results in adding the mapping of $5 \rightarrow 8$ and pages 5, 6, and 7 marked as free space. 

If there is a failure during an insert operation, recovery to a consistent state is possible as the previous page data is still in storage. If the failure occurs after the leaf node is written to storage and no changes are required on the interior nodes above, then recovering the virtual mapping table is sufficient to be in a consistent state.  If the leaf node has been updated on storage, but any splits to interior nodes were not present on storage, recover by re-executing the insert of the key. The key that was inserted is identified by comparing with the previous node version.

\subsection{Implementation Variants}

The B-tree optimizations are deployed as several implementation variants. The base B-tree implementation performs update-in-place writing. This implementation is only executable for storage devices that support a file system interface (with an underlying FTL) or raw flash memory that supports page-level erase-then-write (such as DataFlash \cite{dataflash}). The virtual mapping tree, VMTree, uses a virtual mapping table to minimize write amplification and performs sequential writes to storage. When a logical page is overwritten, the updated page is written to the next sequential page location, and a mapping is inserted in the mapping table to prevent modifying the parent page. The VMTree-OW implementation performs physical page overwrites on memory storage that supports the functionality. A virtual mapping table is not required. 

All implementations use a common page buffer allowing buffering commonly used pages using a LRU algorithm. The root page is always buffered. Each implementation may also utilize a write buffer allowing batching of operations.

\begin{figure}[thbp]
	\centering
	\begin{subfigure}{.47\textwidth}
		\includegraphics[width=\textwidth]{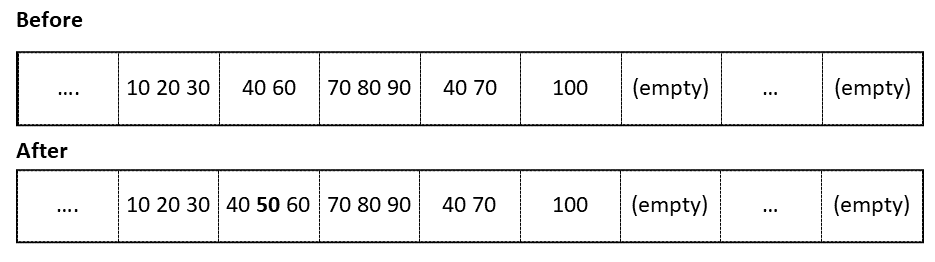}
		\caption{B-tree}
	\end{subfigure}
        \begin{subfigure}{.47\textwidth}
		\includegraphics[width=\textwidth]{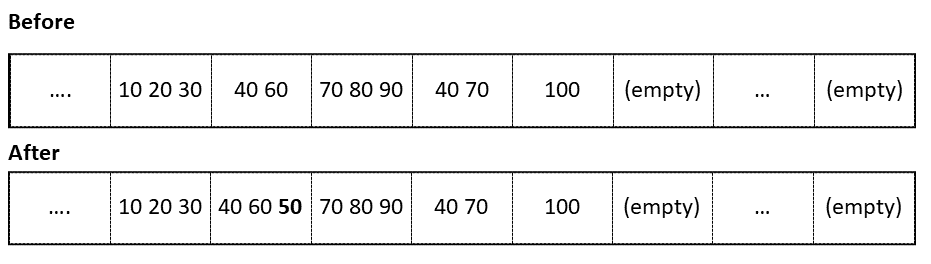}
		\caption{VMTree-OW}
	\end{subfigure}
	\begin{subfigure}{.47\textwidth}
		\includegraphics[width=\textwidth]{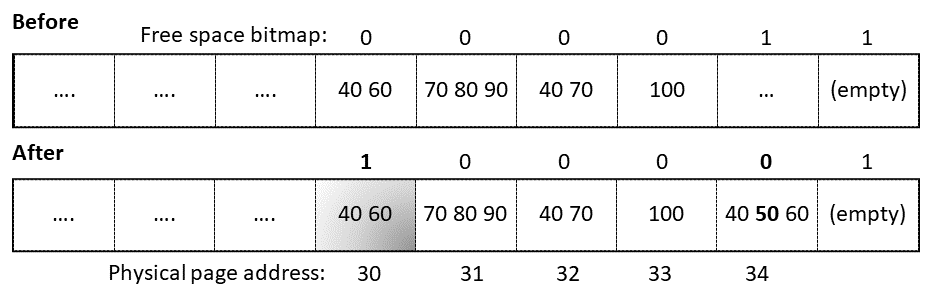}
		\caption{VMTree}
	\end{subfigure}
	\caption{Inserting Key 50}
	\label{fig:insert_example}
\end{figure}

Figure \ref{fig:insert_example} shows how the different variants handle inserting key 50 into the tree in Figure \ref{btree}. The B-tree does not manage free space or memory storage as its relies on the storage device for that. The B-tree performs in-place writing of logical pages, and the physical location can be anywhere on storage as determined by the FTL. The diagram shows adding key 50 to the existing logical page. The B-tree has no control over the physical page location.

VMTree-OW performs overwriting on physical pages. Key 50 is inserted into the existing physical page in the next location. The count bit for the location is set to 0 indicating a valid record is present at that location and valid bit remains at 1. This overwrite guarantees all bit changes on the page are from 1 to 0.

VMTree manages its free space and physical storage. 
Since VMTree writes sequentially and does not overwrite any pages, it will advance through the storage address space faster than the other two algorithms. The free space bitmap is used to determine if a page is used (0) or available (1). Adding the key 50 results in a new page containing (40, 50, 60) being stored at the current write point in storage at physical page address 34. The bit vector for the previous page at address 30 is set to 1 indicating the page can be re-used. A mapping $30 \rightarrow 34$ is added to the mapping table so the parent at address 32 does not need updating.

\section{Experimental Results}

The B-tree implementations were compared on several hardware platforms for multiple data sets. The hardware platforms and storage devices are in Table \ref{expdevices}. The data set characteristics are in Table \ref{datasets}. The results are the average of 3 runs.

\begin{figure*}[ht!]
	\centering
	\begin{tabular}{|l|c|c|c|c|c|}
	\hline
	{\bf Device} & {\bf Storage} & {\bf Page Size} & {\bf R (KB/s)} & {\bf W (KB/s)} & {\bf R/W Ratio} \\
	\hline
    32-bit SAMD21 & SD card & 512 bytes & 500 ; 400 & 500 ; 215 & 1 ; 1.9 \\
    32-bit SAMD21 & DataFlash & 512 bytes & 475 & 35 & 13.5 \\
    16-bit PIC & NAND & 2048 bytes& 203& 187 & 1.1 \\
    16-bit PIC & SD card & 2048 bytes& 198 ; 194 & 108 ; 93 &1.8 ; 2 \\
	\hline
	\end{tabular}
	\caption{Example Embedded Devices and Memory Configurations}
	\label{expdevices}   	
\end{figure*}

\begin{figure*}[tbhp]
	\centering
	\begin{tabular}{|l|c|c|p{6.5cm}|}
	\hline
	{\bf Data Set} & {\bf Key Size}  & {\bf Record Size} & {\bf Description}  \\
	\hline
    Random & 4 bytes & 16 bytes & Random 32-bit integers \\
    Environmental & 8 bytes & 8 bytes & Time series data from \cite{microhash}. Indexing data value (temperature, pressure, or wind speed). \\
    Health & 8 bytes & 8 bytes & WESAD time series data \cite{healthdata} monitoring patient heart, temperature, movement. \\
	\hline
	\end{tabular}
	\caption{Experimental Data Sets}
	\label{datasets}   	
\end{figure*}

 In Figure \ref{expdevices} are each device's read and write performance with the first value being sequential performance and the second being random performance. SD cards have slower random versus sequential performance that varies between cards.  The raw NAND and DataFlash chips have the same performance for sequential and random I/Os. The DataFlash chip allows a page-level write with an erase without erasing a block of pages. This operation is more expensive than writing over an already erased page. The write bandwidth of 35 KB/sec. is for the erase-then-write operation, which is required for the B-tree implementation that performs in-place writes. Storage is pre-erased before an experiment begins.

 The buffer memory allocated to the implementations was M=3 page buffers. The VMTree algorithm also requires additional memory to store the mapping table. The 32-bit SAMD21 platform used page sizes of 512 bytes, while the 16-bit platform had page sizes of 2048 bytes. On the SAMD21 platform, total memory consumed was 3141 bytes consisting of 3 x 512 byte pages, state variables of 330 bytes, bit vector of 251 bytes, and a mapping table of 1024 bytes. On the PIC platform, total memory consumed was 8866 bytes consisting of 3 x 2048 byte buffer pages, state variables of 547 bytes, bit vector of 127 bytes, and a mapping table of 2048 bytes. The B-tree and VMTree-OW did not require the memory for the mapping table or free space bit vector. Additional memory is optionally consumed when using the write buffer.
 
\subsection{Random Data}

The random data set consisted of 10,000 records of 16 bytes with a 4-byte random integer key. The experiment inserted 10,000 records and then performed 10,000 queries. On the SD card, the VMTree time and I/O performance is within 3\% of the B-tree implementation. There is no advantage to maintaining a virtual mapping table as the SD card has its own FTL. The VMTree performs sequential I/O versus random I/O by the B-tree, however, this benefit was not noticeable in the results with the SD card random I/O only about two times slower. VMTree-OW is 8\% slower on the SD card which does not support hardware-level overwrites, as its unsorted leaf structure requires more space and time to process. On the DataFlash, VMTree is about 2\% slower than the B-tree. This memory supports hardware-level overwrites, which results in a dramatic four times performance improvement for VMTree-OW with its overwrite-friendly page structure.

\begin{figure}
    \centering
    \begin{subfigure}{0.5\linewidth}
        \centering
        \includegraphics[width=1\linewidth]{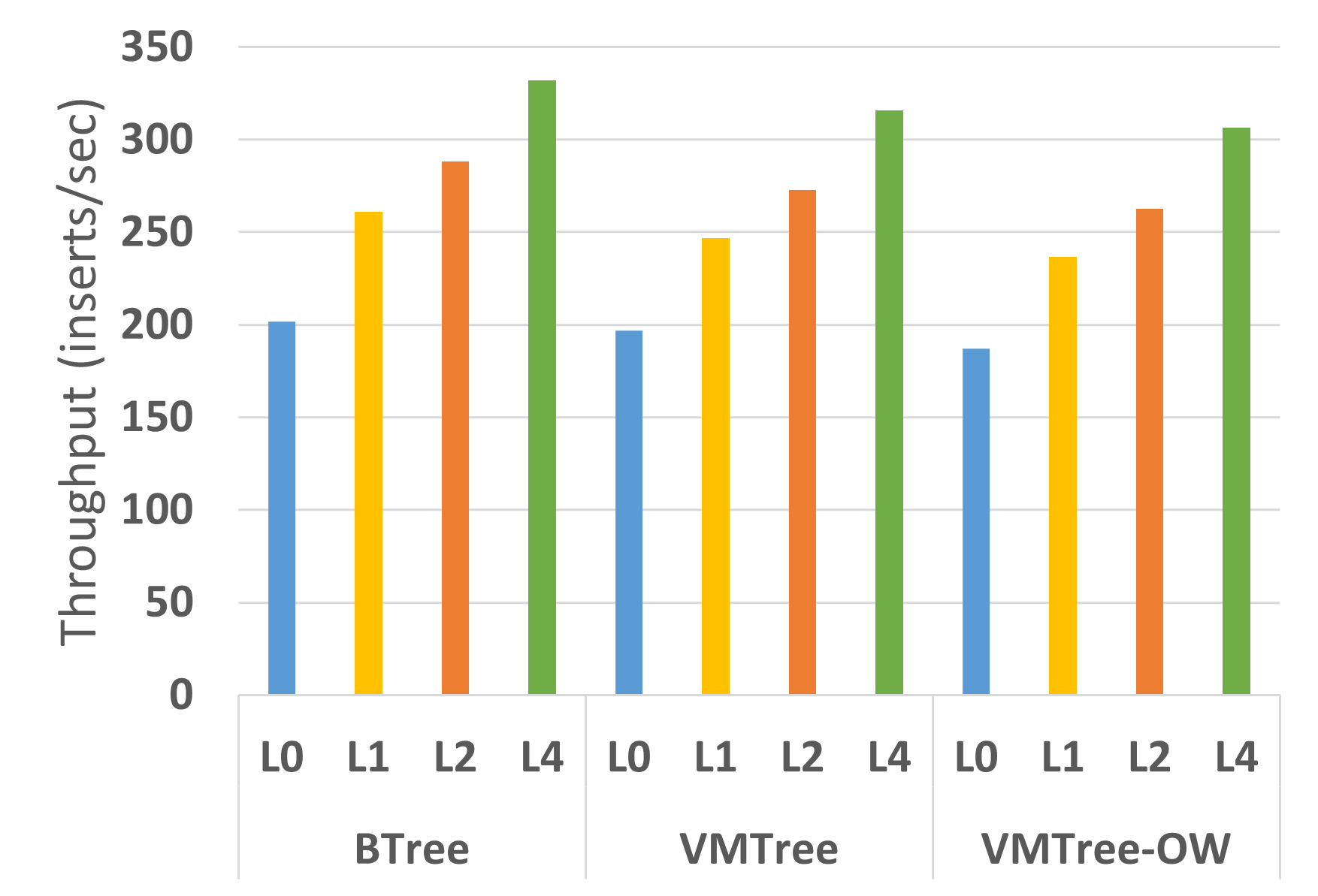}
        \caption{SD Card}        
    \end{subfigure}%
    \begin{subfigure}{0.5\linewidth}
        \centering
        \includegraphics[width=1\linewidth]{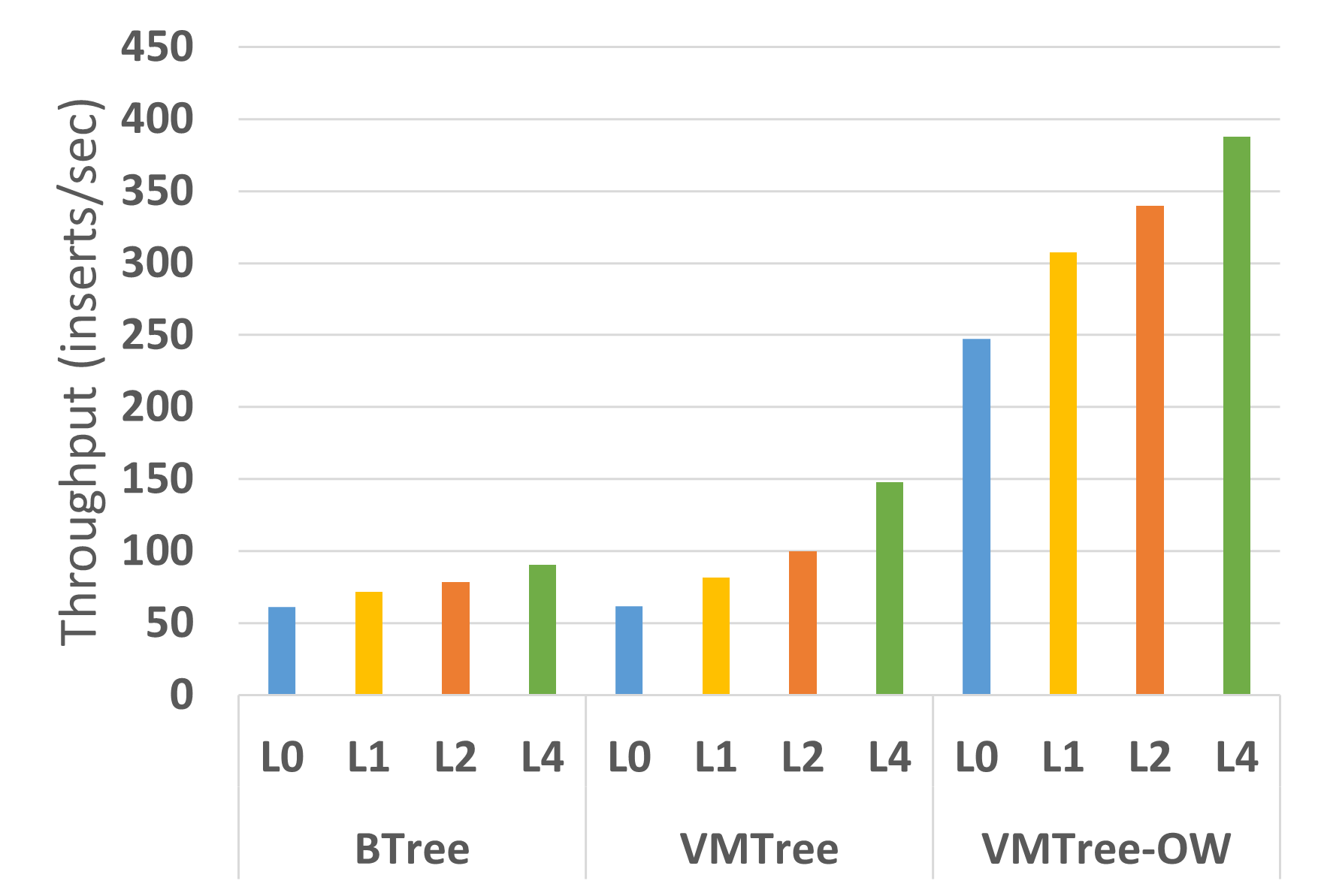}
        \caption{DataFlash}        
    \end{subfigure}
    \caption{32-bit ARM Insert Throughput}
    \label{fig:random_insert}
\end{figure}

Figure \ref{fig:random_insert} shows the insert throughput for each B-tree variant at various write buffer sizes (e.g. {\em L1} is 1 write buffer). The introduction of a write buffer improves all implementations. The benefits of the write buffer are the same as for batch index insertion \cite{batchupdates} with a significant reduction in I/O and overall time. A larger write buffer increases the opportunities to batch writes to the same leaf node resulting in more I/O and time reduction. There is a slightly higher benefit for the VMTree as batching operations reduce the number of virtual mapping lookups during tree traversal. For queries, the read I/Os is identical for all implementations. The query time is within 2\% for VMTree and B-tree. VMTree-OW is 10 to 20\% slower due to its unsorted nodes requiring linear search.

\begin{figure}
    \centering
    \includegraphics[width=0.8\linewidth]{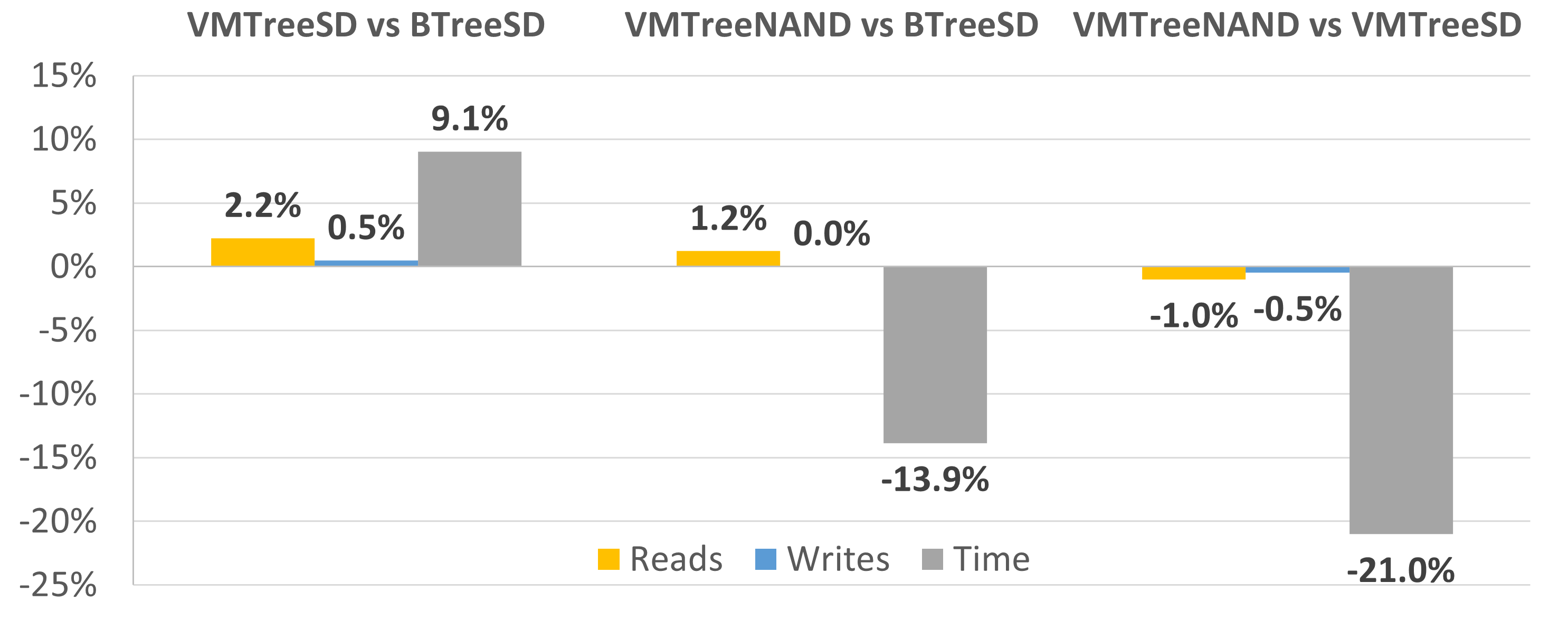}
    \caption{Random Insert Performance on 16-bit PIC on SD Card and NAND}
    \label{primary_insert_vmtree_pic}
\end{figure}

The results on the smaller 16-bit PIC device are similar. Figure \ref{primary_insert_vmtree_pic} displays the relative random insert performance on a SD card and NAND for the B-tree and VMTree. The VMTree-OW is not tested as the NAND memory does not support hardware overwriting, and the B-tree also does not run on the NAND memory, which does not support in-place writing on individual pages. For the pair-wise comparisons in the figure, the percent difference is calculated by ($(tree1 - tree2)/tree2)$. For the VMTreeSD vs BTreeSD, this means VMTreeSD performs 2.2\% more reads, 0.5\% more writes, and takes 9.1\% longer in time. 

The VMTree is slightly slower than the B-tree on the SD card due to some additional I/Os and overhead due to virtual mappings and space management. The VMTree on NAND outperforms both the B-tree and VMTree running on the SD card. It performs the same number of I/Os but executes on a chip with higher write bandwidth. These results demonstrate that the VMTree implementation can run on raw memory with very high efficiency. This is a significant benefit as the NAND chip cost of \$2 to \$4 is 5 times less expensive than an SD card (\$10 to \$20) and easier to deploy on small devices as it is directly on the board. The results demonstrate the advantage of customizing B-tree implementations for storage-specific properties. The addition of a write buffer has substantial performance improvements while not requiring a lot of memory.

\subsection{Sensor Data}

Two sensor data sets were evaluated: environmental weather station data \cite{sbits,microhash} with hourly samples of temperature, precipitation, humidity, and wind speed and the WESAD (wearable stress and affect detection) health data set \cite{healthdata} with sensor measurements including electrocardiogram (ECG), respiration, body temperature, and three-axis acceleration. The B-tree index stores 8-byte index entries consisting of data value (i.e. temperature or ECG) and the record id for retrieval in the time series data set. Although the data is ordered by timestamp, it is not ordered by data value. Data values may have considerable duplication and temporal clustering. The B-tree is a secondary index supporting queries searching on the data value. The buffer memory allocated was M=3 page buffers. The experiment inserted 10,000 records and then ran 10,000 queries on the B-tree.

\begin{figure}
    \centering
    \begin{subfigure}{0.5\linewidth}
        \centering
        \includegraphics[width=1\linewidth]{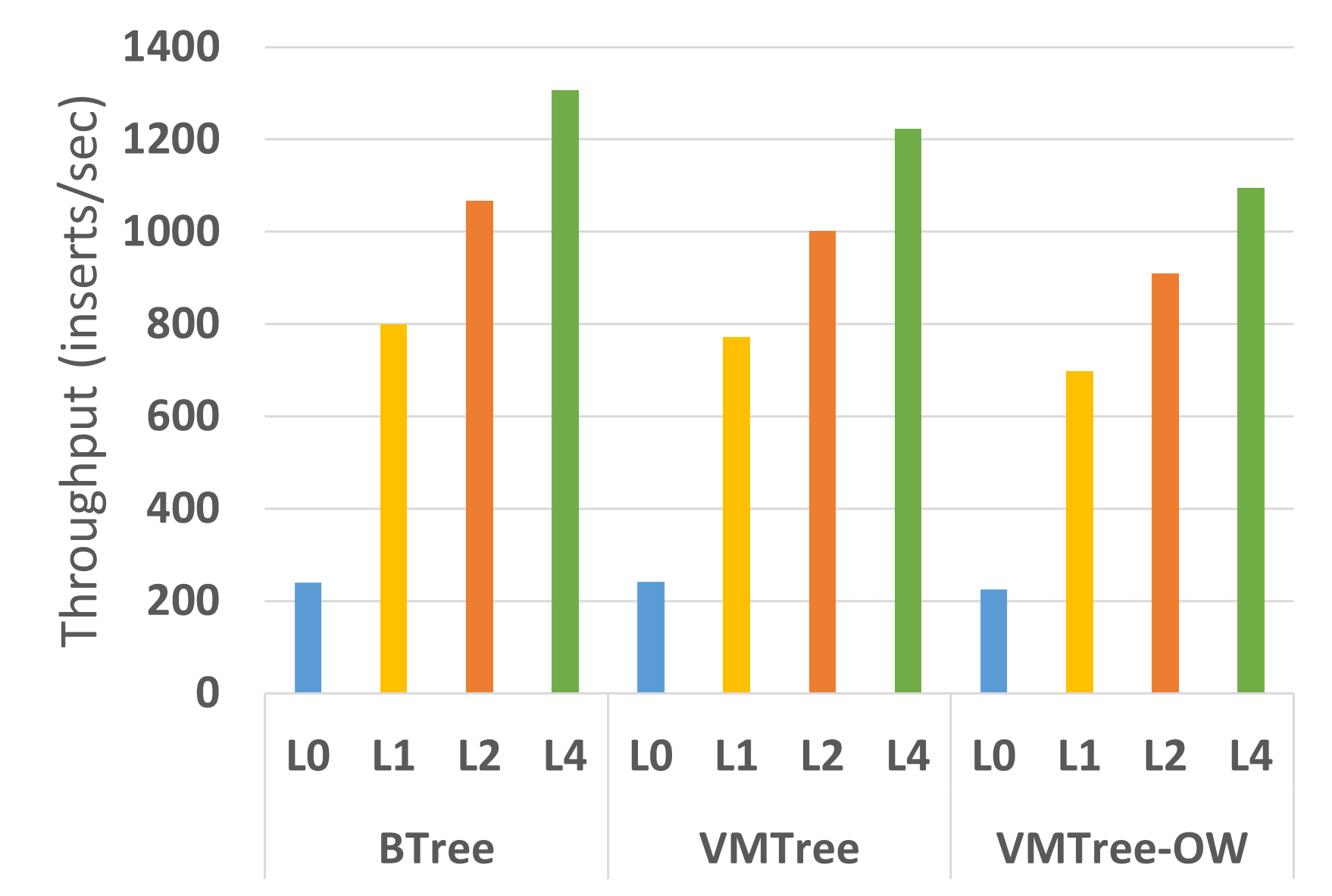}
        \caption{SD Card}        
    \end{subfigure}%
    \begin{subfigure}{0.5\linewidth}
        \centering
        \includegraphics[width=1\linewidth]{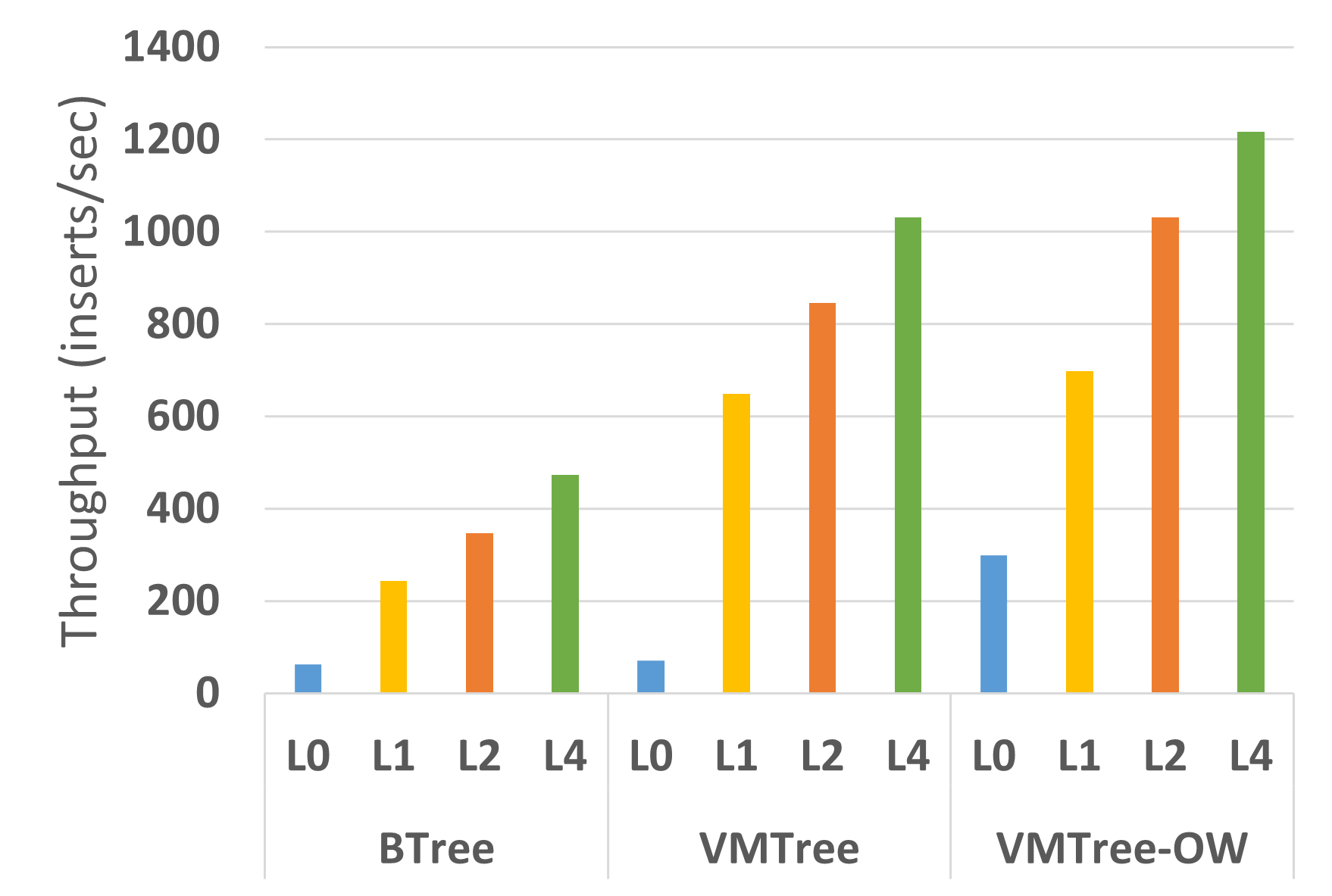}
        \caption{DataFlash}        
    \end{subfigure}
    \caption{Insert Throughput Temperature Data (ARM)}
    \label{fig:temp_throughput}
\end{figure}

The insert throughput results on the 32-bit ARM processor for SD card storage and DataFlash are in Figure \ref{fig:temp_throughput}. The relative performance of the implementations on the SD card is the same as the random data set with VMTree slightly slower than the B-tree. On DataFlash, VMTree-OW has a four times speedup over the B-tree. VMTree is faster than the B-tree as its sequential write pattern over the pre-erased storage is more efficient than the in-place erase-then-write done by the B-tree. A write buffer has a larger benefit compared to random data due to clustering. The temperature data is slowly changing, and batching inserts with one write buffer reduces I/O and time by 63-72\% for all B-tree variants. 

The health data and other environmental variables including humidity, pressure, and wind speed displayed similar performance characteristics. Although the percentage I/O savings varies depending on the amount of duplication and temporal clustering, all sensor data demonstrated substantial savings when using the write buffer. The relative performance of the B-tree variants was consistent regardless of the data set and write buffer size. There were no query differences in any data sets and write buffer sizes as the number of I/Os were identical.

\begin{figure}
    \centering
    \begin{subfigure}{0.5\linewidth}
        \centering
        \includegraphics[width=1\linewidth]{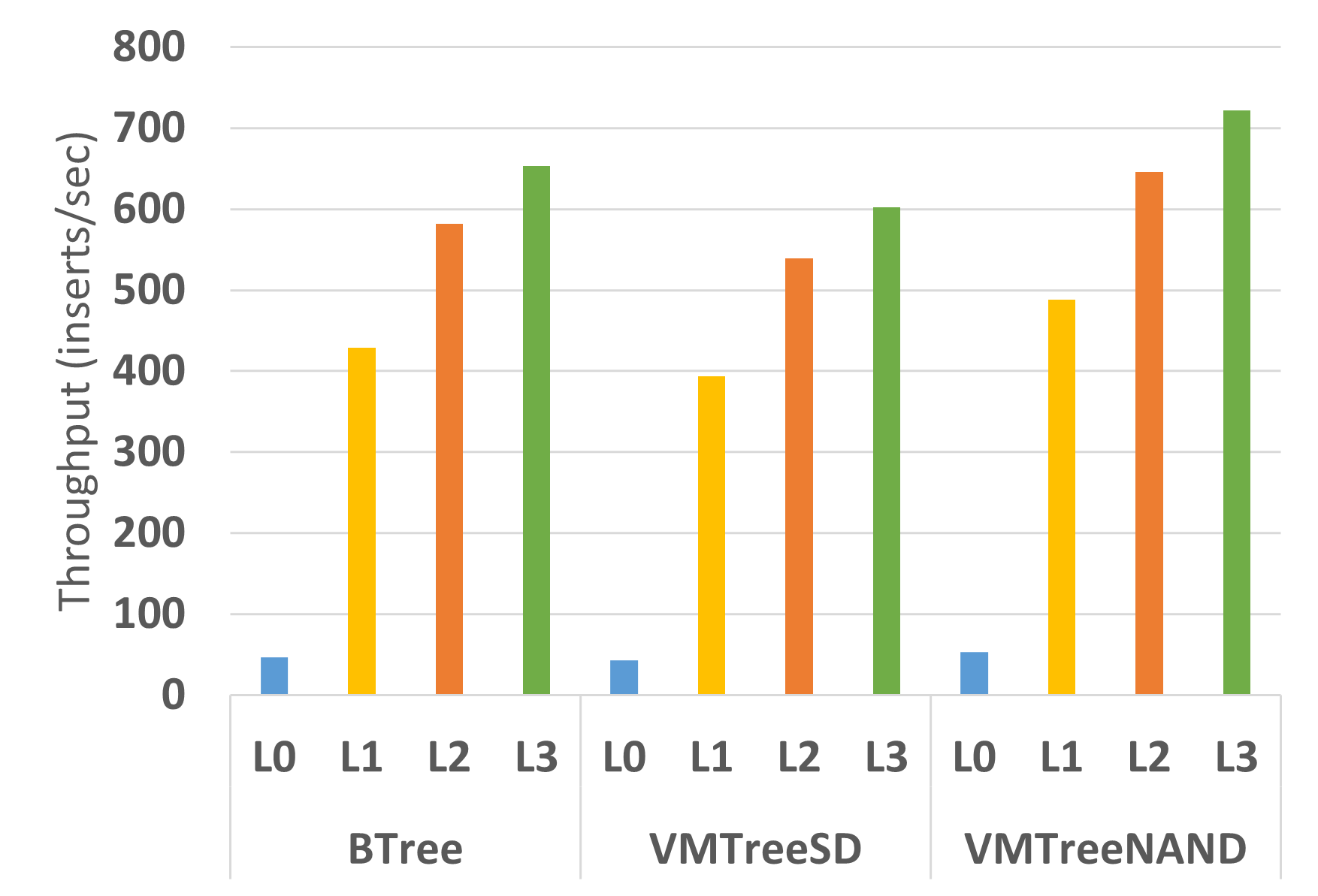}
        \caption{Temperature Data}        
    \end{subfigure}%
    \begin{subfigure}{0.5\linewidth}
        \centering
        \includegraphics[width=1\linewidth]{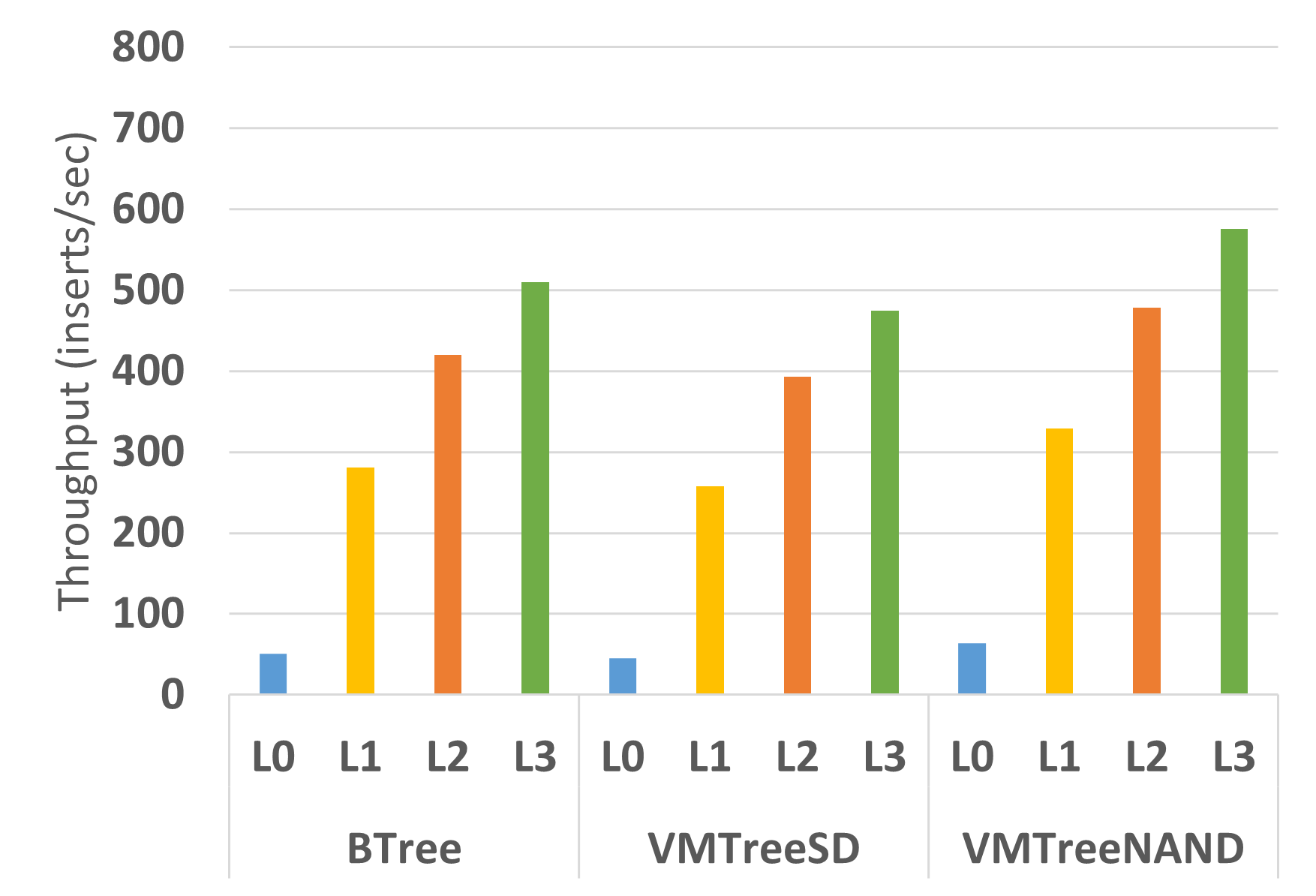}
        \caption{Health Data}        
    \end{subfigure}
    \caption{16-bit PIC Insert Throughput}
    \label{fig:sensor_data_pic}
\end{figure}

On the 16-bit PIC platform, the results are consistent with the 32-bit ARM platform. The insert throughput on the SD card is shown for the temperature and health data sets in Figure \ref{fig:sensor_data_pic}. Addition of one write buffer results in a factor of nine throughput improvement for the temperature data and five times for the health data.  VMTree on NAND consistently has the highest performance.

\subsection{Optimizing Memory Usage}

\begin{figure}[htbp]
    \centering
    \begin{subfigure}{0.5\linewidth}
        \centering
        \includegraphics[width=1\linewidth]{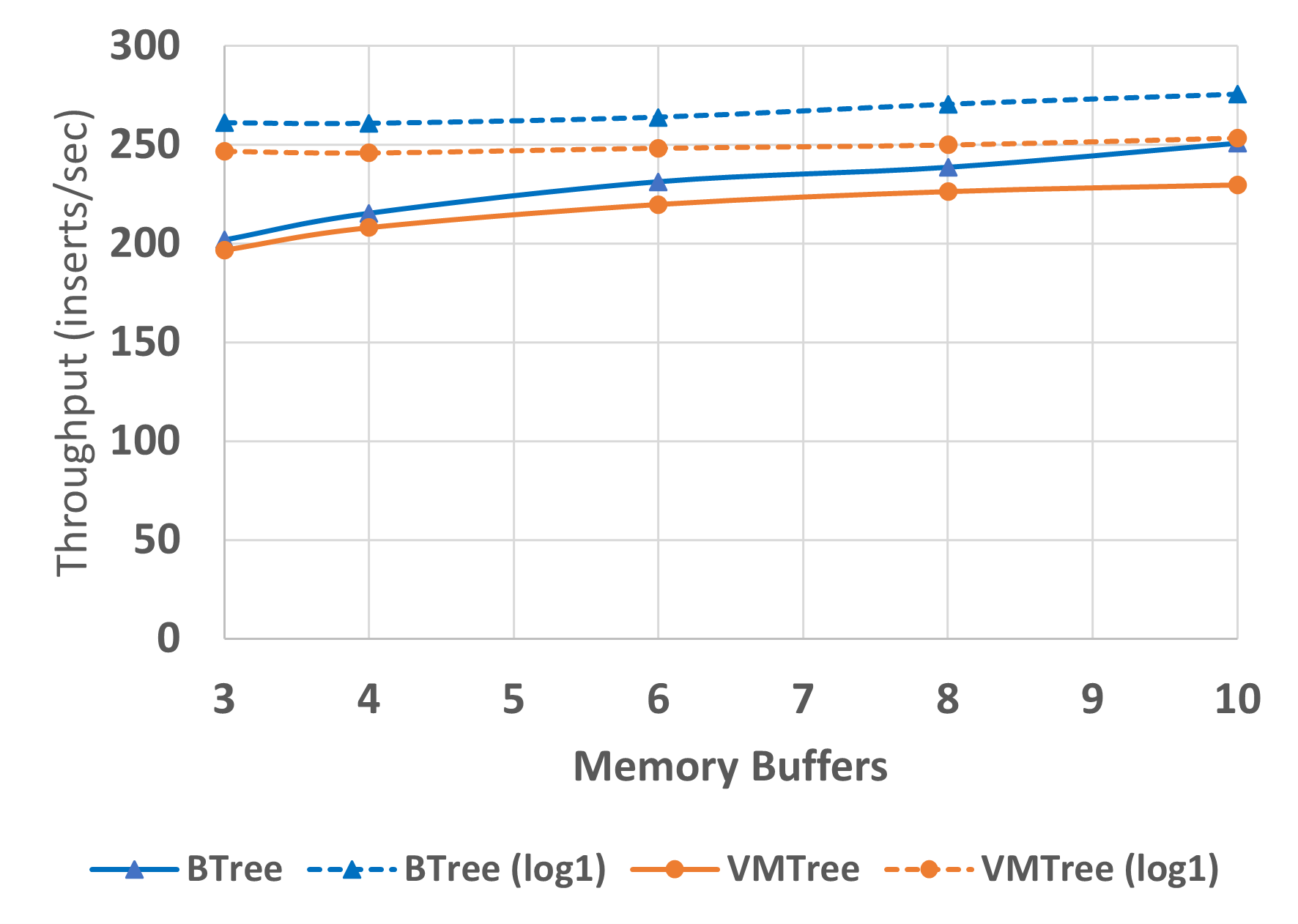}
        \caption{Random Data}        
    \end{subfigure}%
    \begin{subfigure}{0.5\linewidth}
        \centering
        \includegraphics[width=1\linewidth]{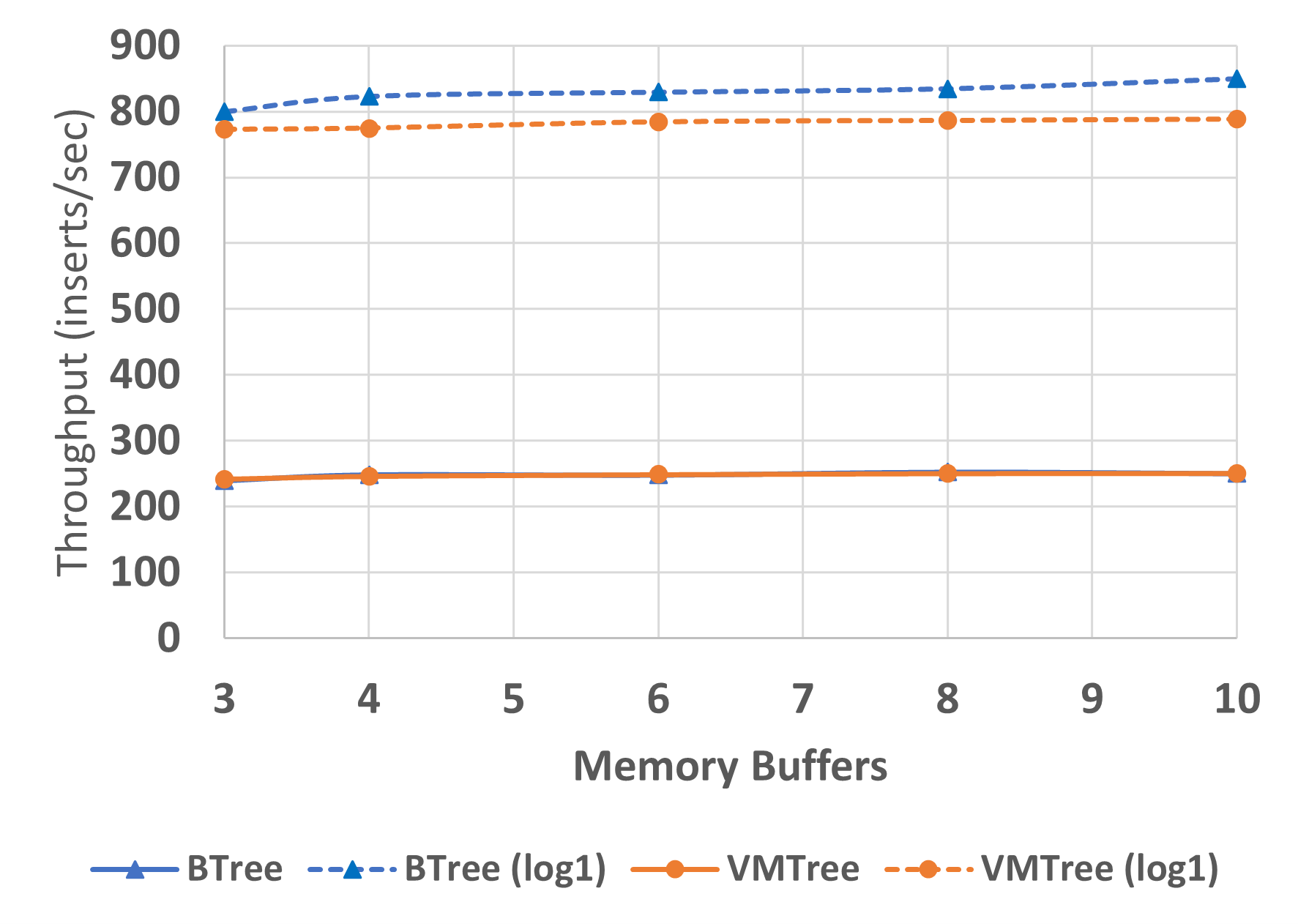}
        \caption{Temperature Data}        
    \end{subfigure}
    \caption{Write Throughput Varying Memory}
    \label{fig:mem_write}
\end{figure}

\begin{figure}[htbp]
    \centering
    \begin{subfigure}{0.5\linewidth}
        \centering
        \includegraphics[width=1\linewidth]{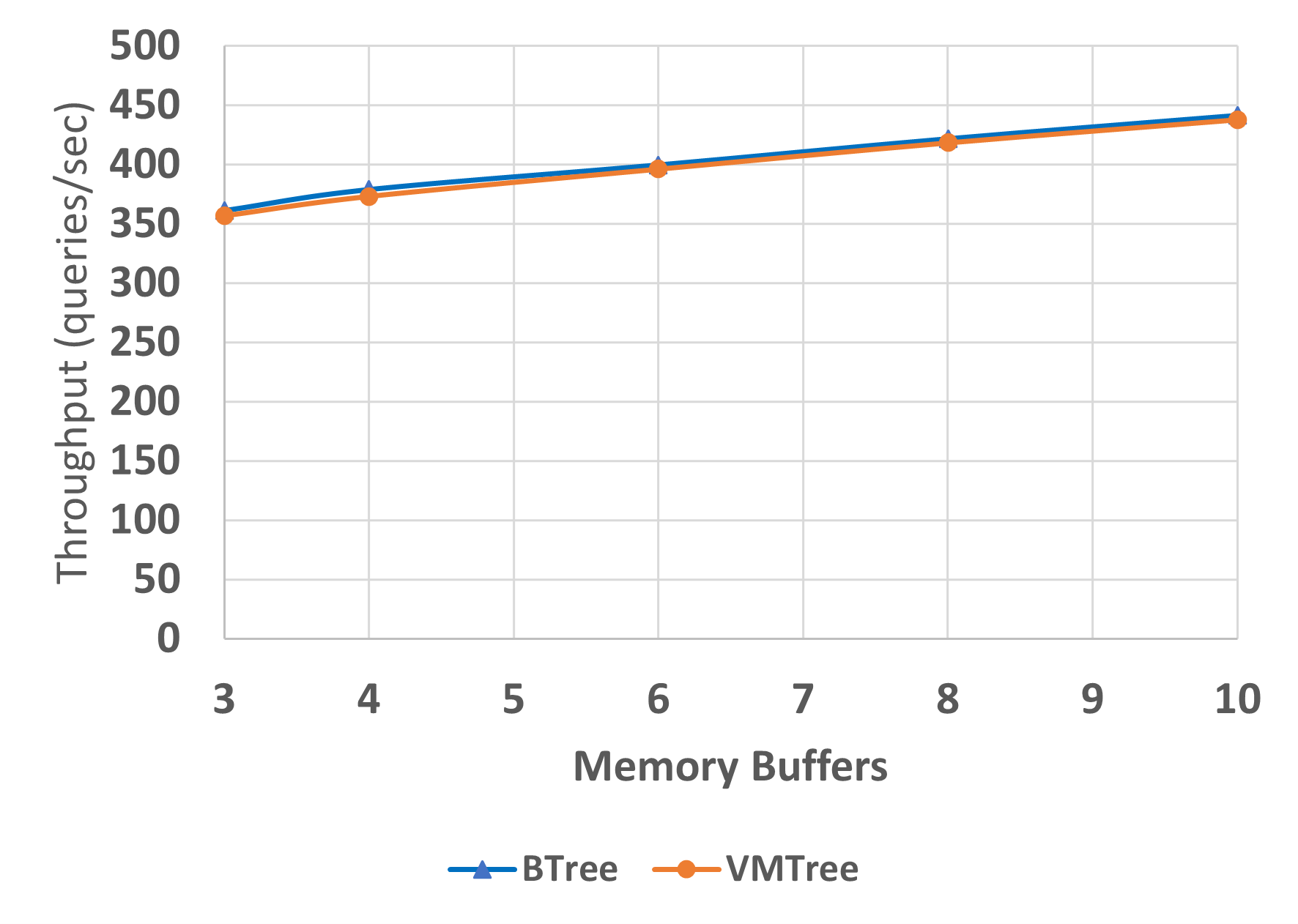}
        \caption{Random Data}        
    \end{subfigure}%
    \begin{subfigure}{0.5\linewidth}
        \centering
        \includegraphics[width=1\linewidth]{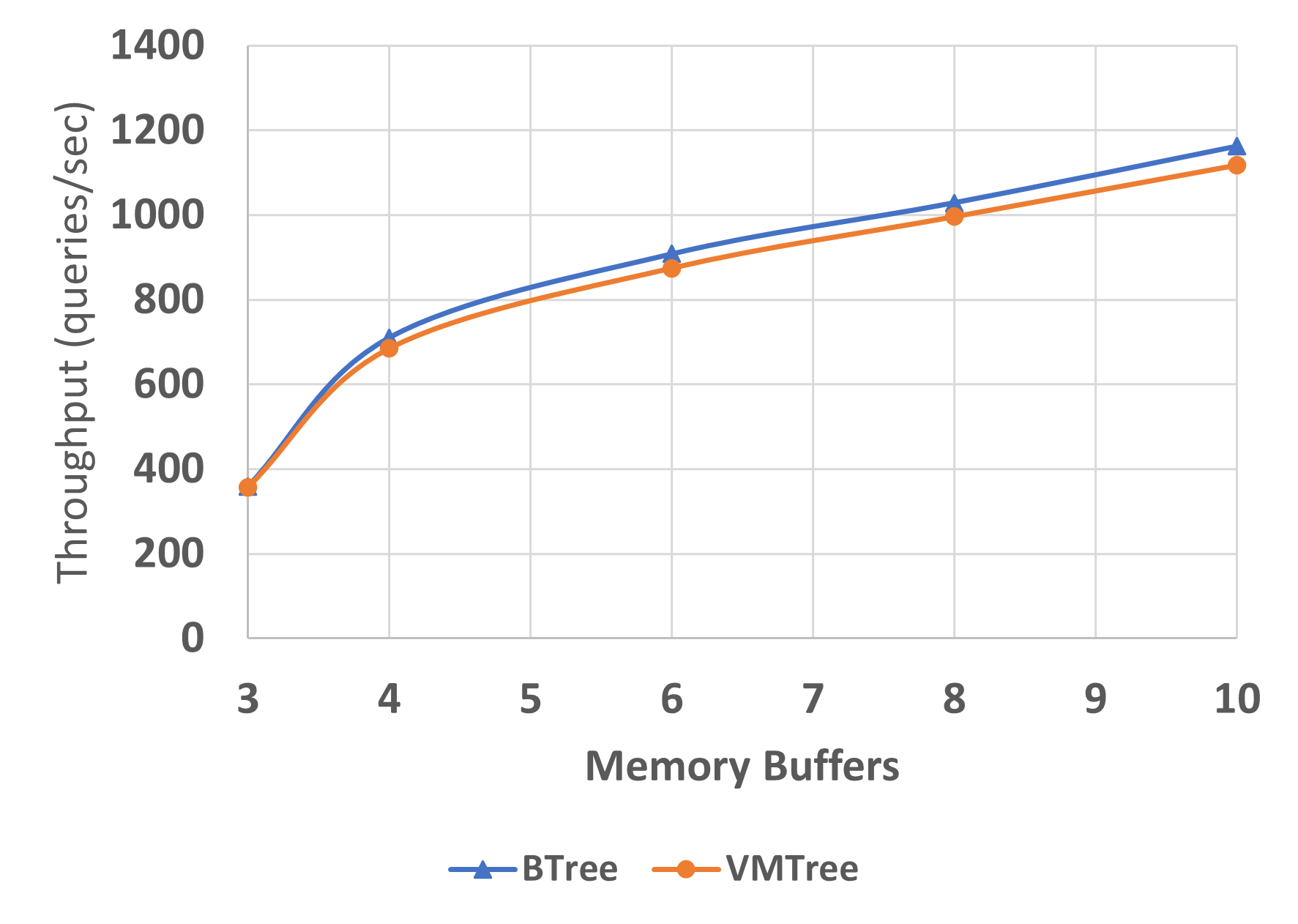}
        \caption{Temperature Data}        
    \end{subfigure}
    \caption{Read Throughput Varying Memory}
    \label{fig:mem_read}
\end{figure}

A key optimization is determining the appropriate way to use any additional RAM. For all B-tree variants, three is the minimum number of buffers to support reads, writes, and tree maintenance. There are two possibilities for allocating additional memory: general page buffers and the write buffer. General page buffers store in RAM pages using an LRU buffer. The write buffer will buffer operations before performing them on the tree and additional pages allow for batching a larger number of operations before modifying the tree. Experiments used the random and temperature data sets, varied the number of page buffers used from 3 to 10, and compared to using a write buffer with 0 pages or 1 page (log1). The results are in Figures \ref{fig:mem_write} and \ref{fig:mem_read} for write and read throughput.

For write throughput, adding page buffers when there is no write buffer increases throughput for the B-tree by 5 to 25\% for random data and less than 5\% for temperature data. The buffer hit rate for temperature data is quite high and does not improve substantially with more buffers. The VMTree is similar except performance improvement for random data is lower between 3 to 14\% due to lower buffer hit rate. Adding more buffers does not change the write I/Os but reduces reads due to more buffer hits. Adding a write buffer has a much larger effect with a significant increase for the temperature data due to clustering. 

Read throughput is not affected by a write buffer. Adding page buffers improves read performance by increasing the buffer hit rate. This is noticeable for data with high duplication and clustering like the temperature data. Going from three to four page buffers allows the system to buffer in memory a complete working path from root to leaf, which avoids having to re-read interior nodes on the path to the leaf when performing splits and tree maintenance.

When utilizing a write buffer to delay and batch updates is feasible, there is more benefit in dedicating memory to the write buffer compared to general page buffers. Sufficient page buffers to buffer a root-to-leaf path is valuable.

\subsection{Varying Data Size}

The VMTree uses more memory than the B-tree and VMTree-OW to maintain its virtual mapping table and free space bitmap. The VMTree also must perform garbage collection when it wraps around in the storage space by moving valid pages before erasing blocks. To quantify these impacts, 100,000 records of random and temperature data was tested. Testing on the SD card allows comparisons with B-tree, which does not run on raw NAND. The storage space was set at 5000 pages, M=4 for page buffers, and the mapping table of 4096 bytes. For 100,000 records, VMTree performed over 135,000 writes requiring 27 passes through storage.

For the random data, VMTree starts off within 1\% of write throughput at 10,000 records. By 100,000 records, B-tree has a throughput 22\% higher. The difference is due to the mapping table being near full resulting in more I/Os. For the temperature data set, the mapping table is only 20\% full, and there is almost no difference in time and I/O between the algorithms. It is the additional I/Os related to collisions in the mapping table that result in slower performance much more than the garbage collection of storage. There is no significant difference in read performance. Both variants scale to larger data sizes. 

\subsection{Varying SD Card Hardware}

SD card performance varies between manufacturers. The performance difference between VMTree and B-tree is impacted based on the ratio of sequential versus random write throughput. VMTree performs sequential writes, which may be higher performing than random writes done by the B-tree. Experiments on the 16 GB SD card with a sequential-to-random write throughput ratio of 2, did not show any significant differences. Five additional SD cards from various manufacturers (Lexar, Sandisk, Kingston) were tested. The sequential-to-random write throughput varied between 1.5 and four times. The impact on VMTree compared to B-tree was less than 5\%. Although older SD cards may have higher sequential-to-random write throughput ratios, the cards tested did not show much difference. This may have been impacted by the slow CPU and bus speed as sequential write performance on all cards was limited by the bus speed not the SD card write speed.

\subsection{Discussion}

The experimental results demonstrate the different performance for the B-tree variants and the advantage of adapting to storage-specific properties. In general, the base B-tree implementation is the best choice for file-based storage where the algorithm does not have to interact with raw physical storage. The VMTree has competitive performance for certain data sizes, but its additional overhead for managing the storage has no benefit unless the storage has a significant difference in random versus sequential write performance. 

VMTree executes on raw NAND flash with high performance and uses only 3 to 4 KB of memory. This allows B-trees to be used on small embedded devices where it was not previously possible. For devices that support page ovewriting, VMTree-OW has significant performance improvements, and unlike VMTree, does not need a virtual mapping table. Write buffers enable batch inserts that have significant performance improvements, especially for sensor data. If the use case is tolerant of data loss of the buffered data, then adding write buffers has the most impact when more memory is available.

\section{Conclusions}

Small embedded devices are key data collectors in Internet of Things applications. Improving data processing and indexing performance on these memory-constrained devices may reduce energy usage and network transmission. This work developed, implemented, and experimentally evaluated several optimized B-tree variants. The  approach of virtual mappings reduces write amplification and allows for efficient B-tree implementations on raw flash memory without requiring a file system interface. The second key optimization was the application of buffering operations as used in prior work on B-trees for servers. When possible for the application use case, write buffering has a dramatic benefit for many sensor data sets with an insert bandwidth 3 to 5 times faster. Optimizing for memory that supports page overwriting also results in a speedup of 4 times. Overall, the three different optimized implementations allow for deployment across a wide-range of devices and memory types and support efficient B-tree indexing while only requiring about 4 KB of memory.

Future work will continue to enhance the optimizations for specific memory types and devices and examine if utilizing other B-tree performance optimizations such as compression have benefit for the sensor data collection environment. The index will be integrated into EmbedDB \cite{embeddb1,embeddb2} that provides a relational and key-value database for embedded systems. The source code for VMTree is at \url{https://github.com/ubco-db/vmtree}.

\bibliographystyle{splncs04}
\bibliography{refs}

\end{document}